\documentstyle[12pt]{article}
\title{{\normalsize
\begin{flushright}SUSX-TH-95/74\\
{\tt hep-ph/9510434}\\
October 1995\\
\end{flushright}}
\vspace{1 cm}
Cosmic Electroweak Strings}
\author{Michael Goodband\thanks{e-mail:
{\tt m.j.goodband@sussex.ac.uk }}
\hspace{1mm} and Mark Hindmarsh\thanks{e-mail:
{\tt m.b.hindmarsh@sussex.ac.uk }}
}
\date{}

\begin{document}
\maketitle
\vspace{-24pt}
\begin{center}
{\it
 School of Mathematical and Physical Sciences\\
University of Sussex\\
Brighton BN1 9QH\\
UK
}
\end{center}

\vfill

\begin{abstract}

We examine the Standard Model field configurations near cosmic strings
in a particular class of models. This class is defined by the condition that
the generator of the flux in the string, $T_s$, commutes with the Standard
Model Lie algebra. We find that if the Standard Model Higgs carries a
charge $F_h /2$ under $T_s$, cosmic string solutions have Z-flux
$\Phi_Z =[n-F_h N/F_{\phi}]4\pi \cos \theta_w /g$, where $n$ is any integer
and $4\pi N/qF_{\phi}$ is the flux of the gauge field associated with $T_s$.
Only the configuration with the smallest value of $|n-F_h N/F_{\phi}|$ is
stable, however. We argue that the instabilities found at higher $\Phi_Z$
are just associated with paths in configuration space reducing
$|n-F_h N/F_{\phi}|$
by one unit. This contradicts recent claims that the instabilities in such
models represent the spontaneous generation of current along the string.
We also show that the stable strings have no Standard Model fermion zero
modes: therefore there is no possibility of supercurrents carried by
Standard Model particles in this class of models.

\end{abstract}
\newpage

In Grand Unified Theories (GUT) of particle physics with spontaneous
symmetry breaking (SSB) there are often topological defects \cite{Vilenkin
and Shellard,Tom and Mark}.
In fact if the symmetries of nature are unified into one
simple Lie group $G$, then there must
exist topological monopoles. The mass density from such monopoles in a
cosmological setting would dominate the universe, and this is not observed.
This is the monopole problem, for which a number of
solutions have been proposed; the leading contenders being inflation or the
formation of strings at a later SSB which connect monopole and
anti-monopole and lead to their annihilation.

The existence of other topological defects, specifically domain walls and
strings, are dependent on the details of the SSBs present in the model.
If domain walls are formed they too will dominate the mass density of the
universe. This can be avoided by inflation or strings carving up the walls.
Such considerations of cosmological implications can lead to restrictions
on the allowable GUTs \cite{Rachel}.

Topological strings on the other hand generally do not lead to cosmological
catastrophies. In fact strings are considered as a possible source of
large scale structure in the universe.

Witten showed \cite{Witten}, that for some particle physics models
strings could be superconducting and support very large currents,
via two different mechanisms. The first involves the
occurrence of a charged scalar condensate, the other involves the
appearance of fermion zero modes on the string. A third mechanism using
charged vector bosons was later identified \cite{Everett,Alford et al}.
All three types depend upon the details of the GUT model and none are
generic.

Recent papers \cite{Peter,P-Davis}, however, have argued that all GUT
scale strings become superconducting at the electroweak symmetry breaking,
and furthermore that a supercurrent spontaneously develops without an
applied electric field. This could have serious cosmological implications,
not least because superconducting string loops can shrink to form stable
rings, and a population of such loops are as disastrous as topological
monopoles.

It is therefore important to check that GUT strings are `generically'
superconducting, and in doing so we return to an old question: how does a
cosmic string affect the fields of the electroweak theory in its vicinity~?

So consider a GUT  with two symmetry breakings and a lagrangian
of the form
$$
{\cal L} = -\frac{1}{4}G_{\mu\nu}^aG^{\mu\nu a} +
\mbox{tr}[(D_{\mu}\phi)^\dagger(D^{\mu}\phi)]
+\mbox{tr}[(D_{\mu}H)^\dagger(D^{\mu}H)]
+V(\phi ,H)
$$
where $D_\mu$ is the covariant derivative, and $V(\phi ,H)$ is some general
gauge invariant fourth order potential.
Now $\phi$ will acquire a vacuum expectation value (vev), which
we take to be at the GUT scale of $10^{16}$ GeV, and breaks the symmetry
group $G$ down  to $G_1$, then $H$ acquires a vev at a lower energy scale
and breaks the symmetry group $G_1$ down further to $G_2$.
In general there can be more stages of symmetry breaking, but we are only
considering two for simplicity. At the first symmetry breaking
some of the various components of the scalar field $H$ will acquire
masses,  while a subset of the $H$ fields will develop an effective
potential that will lead to the second SSB. We will take this
subset to be such that its elements can be identified with the Higgs
doublet of the electroweak standard model, and so the second symmetry
breaking occurs at the electroweak scale of $10^2$ GeV.

Now suppose that the first homotopy group for the first symmetry breaking
is non-trivial,
$\pi_1 (G/G_1) \neq 0$, so that there are stable string solutions
with the asymptotic forms
$$
\phi = \phi_0 \exp(iT_s\theta), \hspace{15mm}
 X^s_{\theta} =T_s/gr
$$
as $r \rightarrow \infty$, where $T_s$ is the string generator.

Now consider the terms that may be present in the lagrangian above
that couple the GUT string fields to the Standard Model fields.
There may be a cross term in the potential $|\phi|^2|h|^2$
between the GUT scalar $\phi$ and the Higgs doublet $h$, and the Standard
Model covariant derivative may contain an additional term proportional to
$X^s_{\mu} h$. These terms were stated in ref.~\cite{P-Davis} to be the
most general terms coupling the GUT string fields to the Standard Model
fields. However, there are a number of other terms that may be present in
the lagrangian above. For example, there could be terms of the form
tr$((\partial_{\mu}{\cal H})X^{s\mu}h)$ and tr$({\cal H}X_{\mu}X^{s\mu}h)$
where ${\cal H}$ is a component of $H$ which is orthogonal to $h$,
$X_{\mu}^s$ is the string gauge field and $X_{\mu}$ is some other vector
boson.
The GUT string may be unstable to solutions with non-zero values of the
fields ${\cal H},h$ and $X_{\mu}$ which could possibly give a charged
condensate on the string. Whether this possibility is realised or not would
depend upon the details of the GUT model.

So we eliminate these later sort of terms from consideration by assuming
that the string generator commutes with all the electroweak generators i.e.
$$
[T_s,\tau^a ]=0 \hspace{5mm} \mbox{and} \hspace{5mm}
[T_s, Y]=0
$$
where $\tau^a$ are the weak isospin generators and $Y$ is the hypercharge
generator. Note that this implies that the GUT string cannot be
superconducting in the sense of Everrett, because $[T_s ,Q]=0$ where $Q$
is the charge generator.

We will further assume that the GUT string is not superconducting in the
sense of Witten either, because the effect of the electroweak phase
transition on a superconducting string has already been considered in
refs.~\cite{Ambjorn et al,Warren}.

To illustrate the effect of the potential term and the addition of
$X^s_{\mu} h$ to the covariant derivative, we will follow
refs.~\cite{Peter,Warren} and extend the electroweak model to include an
extra $U(1)$. So we consider the $SU(2)_L \times U(1)_Y \times U(1)_F$
invariant lagrangian
$$
{\cal L} = -\frac{1}{4}G_{\mu\nu}^a G^{\mu\nu a}
  -\frac{1}{4}B_{\mu\nu}B^{\mu\nu} -(D_{\mu}h)^{\dagger} (D^{\mu}h)
  -\frac{1}{4}X_{\mu\nu}X^{\mu\nu} -(D_{\mu}\phi)^* (D^{\mu}\phi)
  -V(h,\phi)
$$
where the potential is given by
$$
V(h,\phi)=\lambda_h(|h|^2 -\frac{v_h^2}{2})^2 +
 \lambda_{\phi}(|\phi|^2 -\frac{v_{\phi}^2}{2})^2 +
 f(|h|^2 -\frac{v_h^2}{2})(|\phi|^2 -\frac{v_{\phi}^2}{2})
$$
and the covariant derivatives are
\begin{eqnarray}
D_{\mu}\phi &=&(\partial_{\mu}-i\frac{q}{2}F_{\phi}X_{\mu})\phi \nonumber\\
D_{\mu}h &=&(\partial_{\mu}-i\frac{g}{2}\sigma^aW_{\mu}^a
-i\frac{g^{\prime}}{2}B_{\mu}-i\frac{q}{2}F_{h}X_{\mu})h \nonumber
\end{eqnarray}
where $\sigma^a$ are the Pauli spin matrices, $W_{\mu}^a$ are the $SU(2)_L$
gauge fields, $B_{\mu}$ are the $U(1)_Y$ gauge fields and $X_{\mu}$ are the
$U(1)_F$ gauge fields. $h$ is the electroweak Higgs doublet and has a
coupling to the $U(1)_F$ gauge field, while $\phi$ is an electroweak
singlet.

If we assume that this symmetry group is unified into a simple Lie group
$G$ then the $U(1)_F$ charges $F_{\phi}$ and $F_h$ will in general be
rational, but we cannot characterize them further without specifying $G$.

The $U(1)_F$ symmetry is broken first and gives rise to topologically
stable string solutions of the form
$$
\phi = \phi_0 S(r)e^{iN\theta}, \hspace{10mm}
X_{\theta} = \frac{2N}{qF_{\phi}}\frac{A(r)}{r},
$$
where $S(r)$ and $A(r)$ are Nielsen-Olesen profiles \cite{N-O}, and
$N$ is the winding number. The string is taken to be along the z axis.
The scalar field $\phi$ shall be taken as a GUT scale field and so
$\phi_0 \simeq 10^{16}$ GeV, whereas the Higgs field
acquires a vev of the order $10^2$ GeV.
Since the characteristic scale over which a field of mass $m$ varies
is of the order $1/m$, we see
that the characteristic scale of $h$ is fourteen orders of magnitude
bigger than that of $\phi$. So the internal structure of the GUT string is
irrelevant and we need only consider the asymptotic forms of the
Nielsen-Olesen string which are $S(r)=1$ and $A(r)=1$ for
$r \rightarrow \infty$.

We first consider the case $F_h=0$, when the potential term is
the only coupling between the Higgs doublet and the GUT string. The minimum
of the potential is given by
\begin{eqnarray}
|h| & = & 0 \hspace{10mm}\mbox{or}\hspace{10mm}
|h|^2 = \frac{v_h^2}{2}
  -\frac{f}{2\lambda_h}(|\phi|^2 -\frac{v_{\phi}^2}{2}),\nonumber\\
|\phi| & = & 0 \hspace{10mm}\mbox{or}\hspace{10mm}
|\phi|^2 = \frac{v_{\phi}^2}{2}
  -\frac{f}{2\lambda_{\phi}}(|h|^2 -\frac{v_{h}^2}{2}).\nonumber
\end{eqnarray}
The vacuum values are given by $|\phi|^2 = v_{\phi}^{2}/2$ and
$|h|^2 = v_h^{2}/2$ with $v_{\phi}^{2} \gg v_h^{2}$.
If we consider a region where $|\phi|=0$ then the potential energy is
minimized by
$$
|h|^2 = \frac{v_h^2}{2} +\frac{f v_{\phi}^2}{4\lambda_h}.
$$
We can see that for $f>0$ the expectation value of the Higgs is likely to
be raised in the string core, while for $f<0$ it is lowered.
For $f$ sufficiently negative $|h|^2$ can become less than zero so we must
take $|h|=0$. Consequently the electroweak symmetry can be restored about a
GUT string; this is the result given in \cite{Warren}. Note that electroweak
symmetry restoration for $F_h =0$ only occurs for a range of parameters, and
that the above considerations do not give this range because we have ignored
the self-energy potential terms and the kinetic terms.
Conversely, for $f>0$ the electroweak symmetry is always broken in a region
of size $m_h^{-1}$ arround the GUT string \cite{Yajnik}.

We now look for a solution of the form $h^{\dagger} =(0,h_d^{\ast})$ in the
background of the GUT string. Since the GUT string is so massive the
back-reaction of an electroweak field configuration on the GUT string will
be negligible. The equation of motion for $h_d$ is
$$
-\nabla^2 h_d + 2\lambda_h(|h_d|^2 -\frac{v_h^2}{2})h_d
+f(|\phi|^2 -\frac{v_{\phi}^2}{2})h_d=0.
$$
Now for the GUT string $\phi = v_{\phi} S(r) \exp(i\theta)/\sqrt{2}$, where
$S(r)$ is a Nielsen-Olesen profile, and so
$$
f(|\phi|^2 -\frac{v_{\phi}^2}{2}) = \frac{fv_{\phi}^2}{2}(S^2 -1).
$$
The width of the GUT string is approximately
$1/\sqrt{\lambda_{\phi} v_{\phi}^2}$ and so for $v_{\phi} \gg v_h$ the
potential cross term is well approximated by a delta function $\delta(r)$.
For $f$ large and negative the delta function gives the boundary condition
at the origin (taken to be the location of the GUT string),
$h_d^{\prime}(0)=0$. The profile obtained by solving the
equation of motion for $h_d$ with this boundary conditions is shown
in Figure 1. Note that it does not appear to satisfy $h_d^{\prime}(0)=0$.
This is because on the GUT scale the Higgs gradient is given by
$h_d^{\prime}(r_{\phi}) = fm_{\phi} h_d (r_{\phi})/2\lambda_{\phi}$,
and we are considering the limit when $r_{\phi}=1/m_{\phi} \rightarrow 0$.

Essentially, the electroweak symmetry is restored completely on the scale
for which $|\phi|^2 =0$, i.e. on the GUT scale, then $h_d$ returns to its
vacuum value over its characteristic length scale $1/m_h$.

Now when $F_h \neq 0$ the potential term is irrelevant, except possibly for
very large positive values of the parameter $f$. This is because the energy
density  has a contribution of the form $|X_{\theta}h|^2$ from the covariant
derivative, which for the GUT gauge field $X_{\theta} =2N/qF_{\phi}r$ and
the vacuum $h_d = v_h/\sqrt{2}$, will give a logarithmically
divergent contribution to the energy per unit length \cite{Warren,Mark}.

To cancel this logarithmic part of the $\theta $ covariant derivative
$$
D_{\theta}h_d = \left( \frac{1}{r}\frac{\partial}{\partial\theta}
+\frac{ig_z}{2} Z_{\theta} - \frac{iF_h}{F_{\phi}}\frac{N}{r}
\right) h_d,
$$
requires either a $\theta $ dependence for $h_d$, $Z_{\theta} \neq 0$
or both. So consider the field configuration
$$
h_d(r,\theta)=h_d(r)\exp(i\alpha\frac{F_h}{F_{\phi}}\theta), \hspace{10mm}
Z_{\theta} = \frac{\gamma F_h}{F_{\phi}}\frac{2a(r)}{g_zr},
$$
where we take $a(r) \rightarrow 1$ as $r \rightarrow \infty$, and
$\alpha$ such that $\alpha F_h/F_{\phi}$ is an integer, so that $h_d$ is
a single valued function of $\theta$. Substituting these fields into
the covariant derivative above gives
$$
D_{\theta}h_d = \frac{iF_h}{F_{\phi}}\frac{h_d(r,\theta)}{r} (
\alpha + \gamma a(r) - N),
$$
and so to cancel the logarithmic divergence requires $\alpha+\gamma=N$.
Using this we can rewrite the covariant derivative as
$$
D_{\theta}h_d = \frac{i\gamma F_h}{F_{\phi}}\frac{h_d(r,\theta)}{r}
(a(r)-1).
$$
Then the energy of the above configuration is given by
\begin{eqnarray}
E=\pi v_h^2 \int rdr \left[ \left( \frac{dh_d}{dr} \right)^2
+\left( \frac{\gamma F_h}{F_{\phi}} \right)^2 \frac{(a-1)^2}{r^2}h_d^2
+\frac{\beta}{2} (h^2 -1)^2 + \left( \frac{\gamma F_h}{F_{\phi}} \right)^2
\frac{1}{2r^2} \left( \frac{da}{dr} \right)^2 \right]
\end{eqnarray}
where we have rescaled $h_d \rightarrow h_d v_h/\sqrt{2}$,
$Z_{\theta} \rightarrow Z_{\theta} v_h/\sqrt{2}$,
$r \rightarrow r 2\sqrt{2}/g_z v_h$ and
$\beta = 8\lambda_h/g_z^2$, with $g_z^2 = g^2 + g^{{\prime}^2}$.
We are using the standard field basis of
$W_{\mu}^+$, $W_{\mu}^-$, $Z_{\mu}$ and $A_{\mu}$ for the electroweak
fields.
This expression for the energy is the same as for the Nielsen-Olesen
string but with the winding number replaced by $-\gamma F_h/F_{\phi}$,
which is in general non-integer. The profiles $h_d(r)$ and $a(r)$
will therefore be string-like, as can be seen in Figure 2, and the energy
per unit length in the electroweak fields is $\sim \pi v_h^2$.

Electroweak strings are non-topological and it is
possible for them to unwind via `W-condensation' \cite{Achucaro} to
the electroweak vacuum. In the case we are considering it is not possible
for the string-like solution to decay to the electroweak vacuum, because of
the logarithmic term in the energy that would result. There are, however, a
range of possible values of $\alpha$ and $\gamma$ that satisfy the
condition $\alpha + \gamma =N$. To see which values give stable string-like
solutions, we consider $h_u$ and $W_{\mu}^+$ perturbations about the
solution and look for negative modes. Since the GUT string fields are so
massive we need not consider perturbations in the GUT string fields. This
means that the perturbations about the string-like solution above, give
rise to the same perturbation equations as in the electroweak string case
\cite{good} but with
the winding number replaced by $-\gamma F_h/F_{\phi}$. We expand
the perturbations as
$$
\delta h_u = \sum_{m^{\prime}} s_m^{\prime} (r)
e^{im^{\prime}\theta} e^{i\omega t}, \hspace{6mm}
\delta W^+_{\uparrow} = \sum_m -iw_m e^{i(m-1)\theta} e^{i\omega t},
\hspace{6mm}
\delta W^+_{\downarrow} = \sum_m iw_m^{\ast} e^{i(m+1)\theta}
e^{i\omega t}
$$
where $m^{\prime} =m+(\alpha F_h/F_{\phi})$. The symbols $\uparrow$ and
$\downarrow$
refer to the component of spin along the z axis being +1 or -1 respectively.
The resulting perturbation equations in the background gauge
$$
\partial^{\mu} \delta W_{\mu}^+
-ig\cos\theta_W Z^{\mu} \delta W_{\mu}^+
-\frac{ig}{\sqrt{2}} {h_d}^\ast \delta h_u = 0
$$
are
\begin{eqnarray}
\left( \begin{array}{ccc}
       D_1 & A & B \\
       A & D_2 & 0 \\
       B & 0 & D_3
       \end{array} \right)
       \left( \begin{array}{c}
        s_{m^\prime} \\ w_m \\ w_{-m}^{\ast}
        \end{array} \right) & = & \omega^2 \left(
        \begin{array}{c}
         s_{m^\prime} \\ w_m \\ w_{-m}^{\ast}
        \end{array} \right) \nonumber
\end{eqnarray}
where
\begin{eqnarray}
D_1 & = & -\nabla_\rho^2 + \frac{(m^{\prime} -a\epsilon\cos 2\theta_W)^2}
{\rho^2}+\beta (f^2 -1) +2f^2 \cos^2 \theta_W \nonumber \\
D_2 & = & -\nabla_\rho^2 +
\frac{((m-1)-2a\epsilon\cos^2\theta_W)^2}{\rho^2}
+2f^2 \cos^2 \theta_W
- 4\cos^2 \theta_W \frac{\epsilon}{\rho}\frac{da}{d\rho} \nonumber \\
D_3 & = & -\nabla_\rho^2 +
\frac{((m+1) -2a\epsilon\cos^2\theta_W)^2}{\rho^2}
+2f^2 \cos^2 \theta_W
+ 4\cos^2 \theta_W \frac{\epsilon}{\rho}\frac{da}{d\rho} \nonumber \\
A & = & 2\cos\theta_W \left( \nabla_\rho f
+\frac{\epsilon f}{\rho}(1-a) \right) \nonumber \\
B & = & -2\cos\theta_W \left( \nabla_\rho f
-\frac{\epsilon f}{\rho}(1-a) \right) \nonumber
\end{eqnarray}
and we have rescaled as before. The parameter $\epsilon$ is
$\gamma F_h/F_{\phi}$. There are two terms in the perturbation equations
above which can give negative contributions; they are the potential term in
$D_1$ and the last term of $D_2$. This latter term
corresponds to a $-{\bf m.B}$ interaction energy between the
Z-magnetic moment (${\bf m}$) of the W-boson and the Z-magnetic
field (${\bf B}$) of the string-like solution.

We know that for integer values of $\epsilon$ (which will occur for
$F_h/F_{\phi}$ an integer ) the string solution has
negative eigenvalues for modes corresponding to the W-bosons acquiring
non-zero values in the core of the string \cite{good,Margret}. But if the
equations of motion are solved with this `W-condensate', it has been shown
that the solution is gauge equivalent to a string of lower winding
number \cite{Achucaro}. In the case of the string-like solution the
equations of motion for a `W-condensate' are the same as in ref.
\cite{Achucaro} but with the winding number replaced by $-\epsilon$.
Since the generator of the GUT string acts on the
electroweak doublet as a constant times the identity, and the GUT scalar
field is an electroweak singlet, the presence of the GUT string does not
prevent a similar gauge transformation from being made. This will still
be true for non-integer values of $\epsilon$. So if we find negative
modes to the equations above, we must distinguish between those which are
`W-condensation' and those which result in a physical W-boson condensate
trapped in the string core.
The former are unwindings of the string while the later would give a
charged condensate which would break the $U(1)$ of electromagnetism and
so give rise to superconductivity.

If we consider the energy expression (1) we see that
the energy is lower for smaller values of $|\gamma F_h/F_{\phi}|$.
If we consider
$$
\frac{\alpha F_h}{F_{\phi}} + \frac{\gamma F_h}{F_{\phi}} =
\frac{NF_h}{F_{\phi}},
$$
then since $NF_h/F_{\phi}$ is fixed and $\alpha F_h/F_{\phi}$ can only
change by an integer, we conclude that $\gamma F_h/F_{\phi}$ (the $Z$-flux
of the string-like solution in units $4\pi/g_z$) can only change by an
integer. We would expect this lowering of the flux by integer amounts to
occur by `W-condensation' for all $\epsilon$, as it does for integer
$\epsilon$. Thus we expect $\epsilon$ to be lowered by integer amounts
until it lies in the range $-\frac{1}{2}<\epsilon <\frac{1}{2}$.
If the string-like solutions with $\epsilon$ in this range were to have
any negative modes they could not be interpreted as unwindings since
they would raise the energy, and so would have to be interpreted
as the occurrence of a physical charged condensate.

To investigate the above arguments numerically for a GUT string of winding
$N=1$, we consider $F_h/F_{\phi}$ values of
(a) 0, (b) 1, (c) 0.4 and (d) 0.5. Case (c) is actually realised in an
$SO(10)$ model considered by Alford and Wilczek \cite{Alford and Wilczek}.

For case (a) the $\delta$-function potential cross term is the only
coupling between the GUT string and the electroweak fields, and its only
effect is to give $h_d(0)^{\prime} =0$. As seen earlier, this condition is
satisfied on the GUT scale but is negligible on the electroweak scale, and so
for string solutions the effect of the potential term
on the profiles is negligible. So the string solutions and their stability
are the same as for electroweak strings solutions; they are unstable for
physical values of parameters \cite{good,Margret}.
The `vacuum' solution in this case is that shown in Figure 1.

For case (b) $\epsilon$ will be an integer, and so the string solutions and
their stability will again be the same as for the electroweak string. The
`vacuum' in this case will have $\epsilon =0$, but the Higgs field will still
have a winding in order to cancel the logarithmic contribution to the energy
from the GUT gauge field. The `vacuum' solution is again given by the profile
in Figure 1.

For case (c) first consider $(\alpha,\gamma )=(5/2,-3/2)$ which gives
$\epsilon = -0.6$ and is outside our proposed stability range. The profiles
for the string-like configuration were solved for by a relaxation method on
the energy (1) and substituted into the perturbation equations. These were
then solved by direct matrix methods for $\sin^2 \theta_w =0.23$.
A negative mode was found for angular momentum $m=-1$. As with the
electroweak string, this mode is interpreted as an instability to the
winding ($\epsilon$) increasing by one unit.
The stability line i.e.~the line in $(\beta, \theta_w)$ parameter
space for which $\omega^2 =0$, is an approximate vertical line at about
$\theta_w =\pi /4$. For $\epsilon < -0.6$ this line moves up to higher
$\theta_w$, while for $\epsilon > -0.6$ this line moves down to lower
$\theta_w$ and so for some $\epsilon$ we would expect no negative modes
to occur at $\sin^2 \theta_w = 0.23$.

Now consider case (c) with $(\alpha,\gamma )=(0,1)$ which gives
$\epsilon =0.4$, i.e. it is the solution the above configuration decayed
to. This had no negative modes and so is a stable solution.

For case (d) the parameter values $(\alpha,\gamma ) =(0,1)$ and
$(\alpha,\gamma )=(2,-1)$ have $\epsilon$ values of $+0.5$ and $-0.5$
respectively and so the two solutions are degenerate in energy.
For $\sin^2 \theta_w =0.23$ both of these solutions were found to be
stable. For $\sin^2 \theta_w =0$ the $\epsilon =-0.5$ solution was found to
have an $m=-1$ zero mode while the $\epsilon =0.5$ solution had as
$m=1$ zero mode.  Integer $\epsilon$ strings also have zero modes at
$\sin^2 \theta_w =0$, and these also occur at angular momentum
$m=2\epsilon$ \cite{good}. These modes are to be interpreted as transitions
between the $-\epsilon$ and $+\epsilon$ solutions via a W-string.
For $\sin^2 \theta_w >0$ the energy of the W-string is above that of the
corresponding Z-string and so there is a barrier to such transitions, while
for $\sin^2 \theta_w =0$ the W-string and Z-string solutions are degenerate
in energy.

So at $\sin^2 \theta_w =0$ all strings with $|\epsilon| \leq 0.5$ are
stable. At $\sin^2 \theta_w =0.23$, in addition to the stable strings
above there are metastable string solutions for $|\epsilon|$ in the
approximate range $0.51$~--~$0.53$ for $\beta$ in the range
$0.25$~--~$4.0$.

Now in \cite{Peter} it was claimed that because the electroweak symmetry
was restored the W-bosons would be massless, since the W-boson gets its
mass from a term proportional to $|h_d|^2$. The confinement energy was,
however, ignored and since the potential well is approximately $1/m_h$
wide and the characteristic scale of the W-boson $1/m_W$ is comparable,
the confinement energy will be sizeable.

So we looked at the W-boson bound modes at $\sin^2(\theta_w)=0.23$
for $\sqrt{\beta}=0.5,1.0$ and $2.0$ for the various cases above.
The eigenvalues obtained for the bound W-bosons in case (a) with angular
momentum $m=0$, are $\omega =0.88,0.91$ and $0.92$ $m_W$ respectively
($m_W$ is the mass of the W-boson in the vacuum), which in view of the above
comments is to be expected. The remaining cases also possessed bound
W-bosons for angular momentum $m=0$ with similar sized eigenvalues.

For angular momentum $m= -1$ the most likely case to have bound W-bosons
with significantly lower eigenvalues than above is $|\epsilon |=0.5$, since
there are zero modes at $\sin^2 \theta_w =0$.
We found that for $\epsilon =0.5$ the lowest bound mode eigenvalues
are $\omega = 0.25$, $0.27$ and $0.28$ $m_W$ respectively. These are the
lowest bound mode eigenvalues that were obtained for any of the stable strings
for the $\theta_w$ and $\beta$ values given above. We therefore find no
massless W-boson states on the strings \cite{Patrick}.

Finally we consider whether there are any fermion zero modes present
on the stable string-like configurations. We know that electroweak
strings possess fermion zero modes \cite{ESfermzero} and so we might
expect there to be fermion zero modes on the string-like solutions as well.
To investigate the possible existence of
fermion zero modes consider the
$SU(2)_L \times U(1)_Y \times U(1)_F$ invariant lagrangian for the
first family of leptons
$$
{\cal L}_{\rm ferm} = -i\overline{\psi} \gamma^{\mu} D_{\mu}\psi
- ie_R \gamma^{\mu} D_{\mu}e_R
+ Y_e(e_R h^{\dagger}\psi +\overline{\psi} he_R)
$$
where
\begin{eqnarray}
D_{\mu}\psi & = & \left( \partial_{\mu} - \frac{ig\tau^a}{2}W_{\mu}^a
+\frac{ig^{\prime}}{2}B_{\mu} - \frac{iq}{2} F_L^e X_{\mu}
\right) \psi \nonumber \\
D_{\mu} e_R & = & \left(\partial_{\mu} + ig^{\prime} B_{\mu}
-\frac{iq}{2} F_R^e X_{\mu} \right) e_R \nonumber
\end{eqnarray}
and $Y_e$ is a constant. For the Yukawa coupling term to be $U(1)_F$
invariant we must have $F^e_L - F^e_R = F_h$.

The Dirac equations in the presence of the GUT string with an
electroweak string-like configuration about it are
\begin{eqnarray}
(iD_0 + i\sigma_k D_k )e_L & = & mh_de_R \nonumber \\
(iD_0 - i\sigma_k D_k )e_L & = & mh^{\ast}_de_R \nonumber
\end{eqnarray}
where
\begin{eqnarray}
D_k e_L = \left( \partial_k + \frac{ig_z}{2} \cos 2\theta_w Z_k
-\frac{iq}{2}F_L^e X_k \right) e_L \nonumber \\
D_k e_R = \left( \partial_k - ig_z \sin^2 \theta_w Z_k
-\frac{iq}{2}F_R^e X_k \right) e_R \nonumber
\end{eqnarray}
Now we write $e_R^T =(c_1,c_2)$, $e_L^T =(d_1,d_2)$ and expand
$c_1$, $c_2$, $d_1$ and $d_2$ as
\begin{eqnarray}
c_1 & = & c_1(r) \exp (ikz-i\omega t
+i\left( m-\frac{\alpha F_h}{F_{\phi}} \right) \theta ) \nonumber \\
c_2 & = & ic_2(r) \exp (ikz-i\omega t
+i\left( m+1-\frac{\alpha F_h}{F_{\phi}} \right) \theta ) \nonumber \\
d_1 & = & d_1(r)\exp (ikz-i\omega t+im\theta ) \nonumber \\
d_2 & = & id_2(r)\exp (ikz-i\omega t +i(m+1)\theta ) \nonumber
\end{eqnarray}
where $m$ is the angular momentum of the mode. We are looking for
zero modes so we consider $|\omega | =|k|=0$. Then the equations
separate into two pairs of coupled equations,
\begin{eqnarray}
d_1^{\prime} -\frac{d_1}{r} \left[ \left( m-\frac{NF^e_L}{F_{\phi}}
\right) + \frac{\gamma F_h}{F_{\phi}} a(r)\cos 2\theta_w \right]
& = & Y_e h_d(r) c_2 \nonumber \\
c_2^{\prime} + \frac{c_2}{r} \left[ \left(
m+1-\frac{\alpha F_h}{F_{\phi}} -\frac{NF_R^e}{F_{\phi}} \right)
-\frac{\gamma F_h}{F_{\phi}} a(r) 2\sin^2 \theta_w \right]
& = & Y_e h_d(r) d_1
\end{eqnarray}
and
\begin{eqnarray}
c_1^{\prime} -\frac{c_1}{r} \left[ \left(
m-\frac{\alpha F_h}{F_{\phi}} -\frac{NF^e_R}{F_{\phi}}
\right) - \frac{\gamma F_h}{F_{\phi}} a(r)2\sin^2\theta_w \right]
& = & -Y_e h_d(r) d_1 \nonumber \\
d_2^{\prime} + \frac{d_2}{r} \left[ \left(
m+1-\frac{NF_L^e}{F_{\phi}} \right)
+\frac{\gamma F_h}{F_{\phi}} a(r) \cos 2\theta_w \right]
& = & -Y_e h_d(r) c_1
\end{eqnarray}
For there to be a zero mode solution both fields in the pair must
be non-singular at the origin \cite{fermionzero}. For (2) this requires
$$
m-\frac{NF_L^e}{F_{\phi}} \geq 0 \hspace{4mm} \mbox{and} \hspace{4mm}
\frac{\alpha F_h}{F_{\phi}} + \frac{NF_R^e}{F_{\phi}} -m-1 \geq 0.
$$
Using $F_R^e =F_L^e -F_h$ gives
$$
m \geq \frac{NF_L^e}{F_{\phi}} \hspace{4mm} \mbox{and} \hspace{4mm}
(\alpha -N)\frac{F_h}{F_{\phi}} +\frac{NF_L^e}{F_{\phi}} -1 \geq m
$$
and so for both conditions to be true, we must have
$$
\gamma \frac{F_h}{F_{\phi}} \leq -1
$$
where we used $\alpha + \gamma =N$. Similarly for (3) we get
$$
\gamma \frac{F_h}{F_{\phi}} \geq 1.
$$
So for there to be fermion zero modes we must have
$|\gamma F_h/F_{\phi} | \geq 1$ but we showed earlier that the
string-like solutions are only stable for
$|\gamma F_h/F_{\phi} | \leq 1/2$, and so the stable solutions do not have
fermion zero modes.

So in conclusion we find that if the only coupling between the GUT string
fields $(\phi ,X_{\theta})$ and the electroweak fields is a
$qF_h X_{\mu} h/2$ term in the covariant derivative and a $|\phi|^2|h|^2$
term in the potential, then there are electroweak string-like solutions
about the GUT string. The Z-flux of such strings is
$\Phi_Z =(n-F_h N/F_{\phi} )4\pi \cos\theta_w /g$, where $n$ is an integer
and $N$ is the integer winding of the GUT string. We found no evidence for
the formation of stable charged condensates: the strings with
$|n-F_h N/F_{\phi}| > 1/2$ did possess negative modes, but we surmise these
instabilities are due to the string decaying to one with lower Z-flux by the
`W-condensation' mechanism of ref.~\cite{Achucaro}.
Those strings with $|n-F_h N/F_{\phi}| \leq 1/2$ possess no negative modes and
so GUT strings can have stable electroweak strings arround them, similar to
those found arround global strings in a two Higgs doublet model in
ref.~\cite{Dvali}.

The mechanism for superconductivity given in ref. \cite{Peter} required the
occurrence of W-boson zero modes on the string. We have shown that these do
not occur for this class of string solutions and so supercurrents do not
arise as claimed in refs.~\cite{Peter,P-Davis}.

We have further shown that the stable string solutions do not possess fermion
zero modes, and so conclude that a non-superconducting GUT string does not
become superconducting after the electroweak phase transition.

The effects of there being stable electroweak string solutions about GUT
strings should be negligible. First of all, particle production by the string
due to the coupling between the GUT string and light particles has been
considered in ref.~\cite{Srednicki}, where it was shown that gravitational
radiation was a more significant energy loss mechanism. We would not expect
any significant change in the dynamics of the GUT strings due to the forces
between the electroweak strings because they are negligible in comparison to
the GUT string mass.

In ref.~\cite{Linked strings} a baryon production mechanism was outlined
which involved the de-linking of linked electroweak strings. Electroweak
strings are, however, highly unstable and so it is unclear whether or not
they form. Here we have stable electroweak strings forming about GUT strings.
However, the GUT string network will have reached a scaling solution by the
electroweak phase transition and so the
number density of linked strings would be extremely low. The net baryon
number produced by this mechanism would be negligible.

\subsection*{Acknowledgements}
This work was supported by PPARC: MG by studentship number
93300941 and MH by Advanced Fellowship number B/93/AF/1642.

\newpage

\section*{Figure captions}

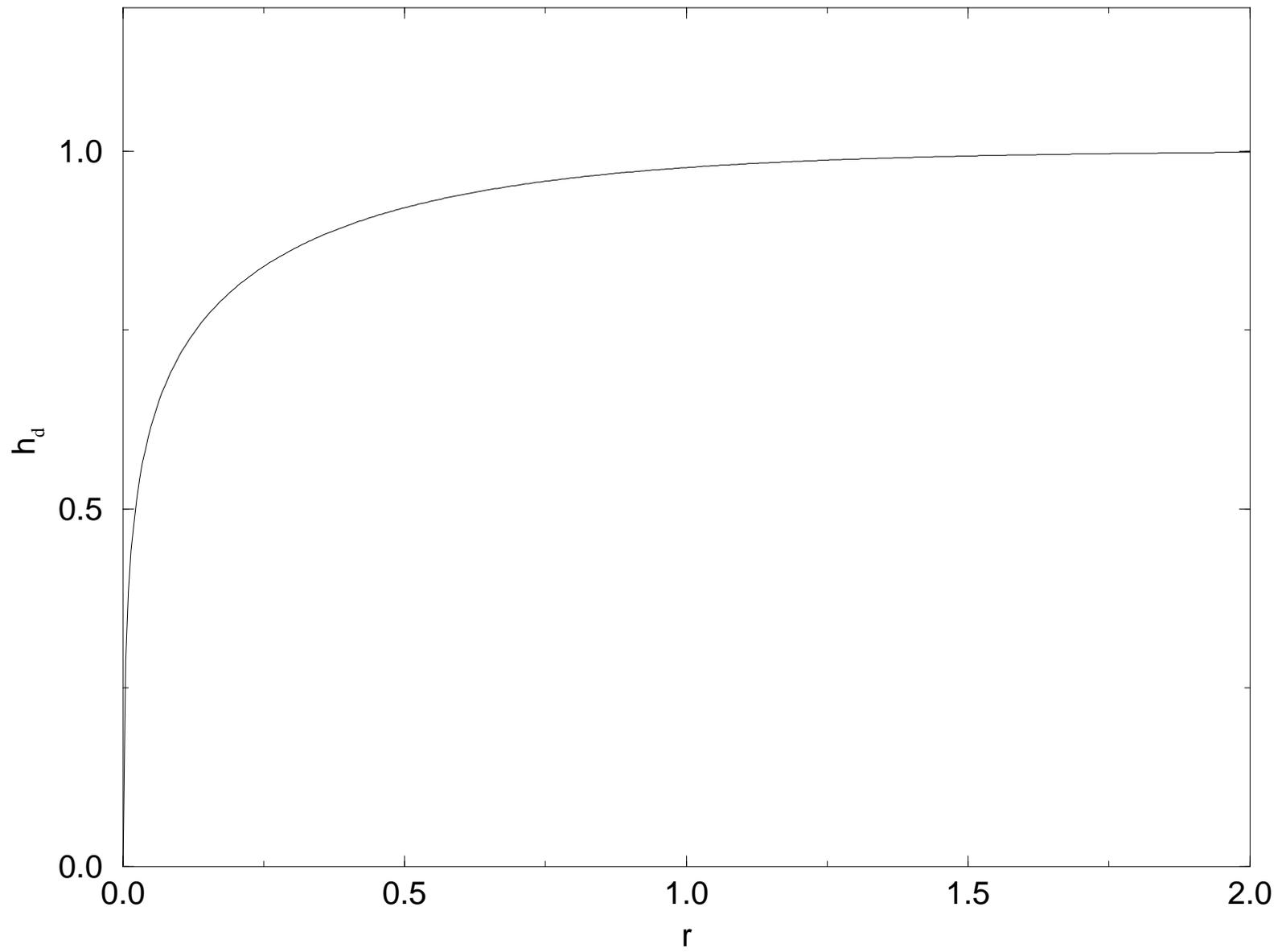
{\bf Figure 1:} $h_d$ profile showing symmetry restoration about a
GUT string for $F_h =0$.
\newline
{\bf Figure 2:} $h_d(r)$ and $a(r)$ profiles for $\epsilon =0.4$
(solid line) and those for a Nielsen-Olesen string  (dashed line)
($\epsilon =1$ for comparison)

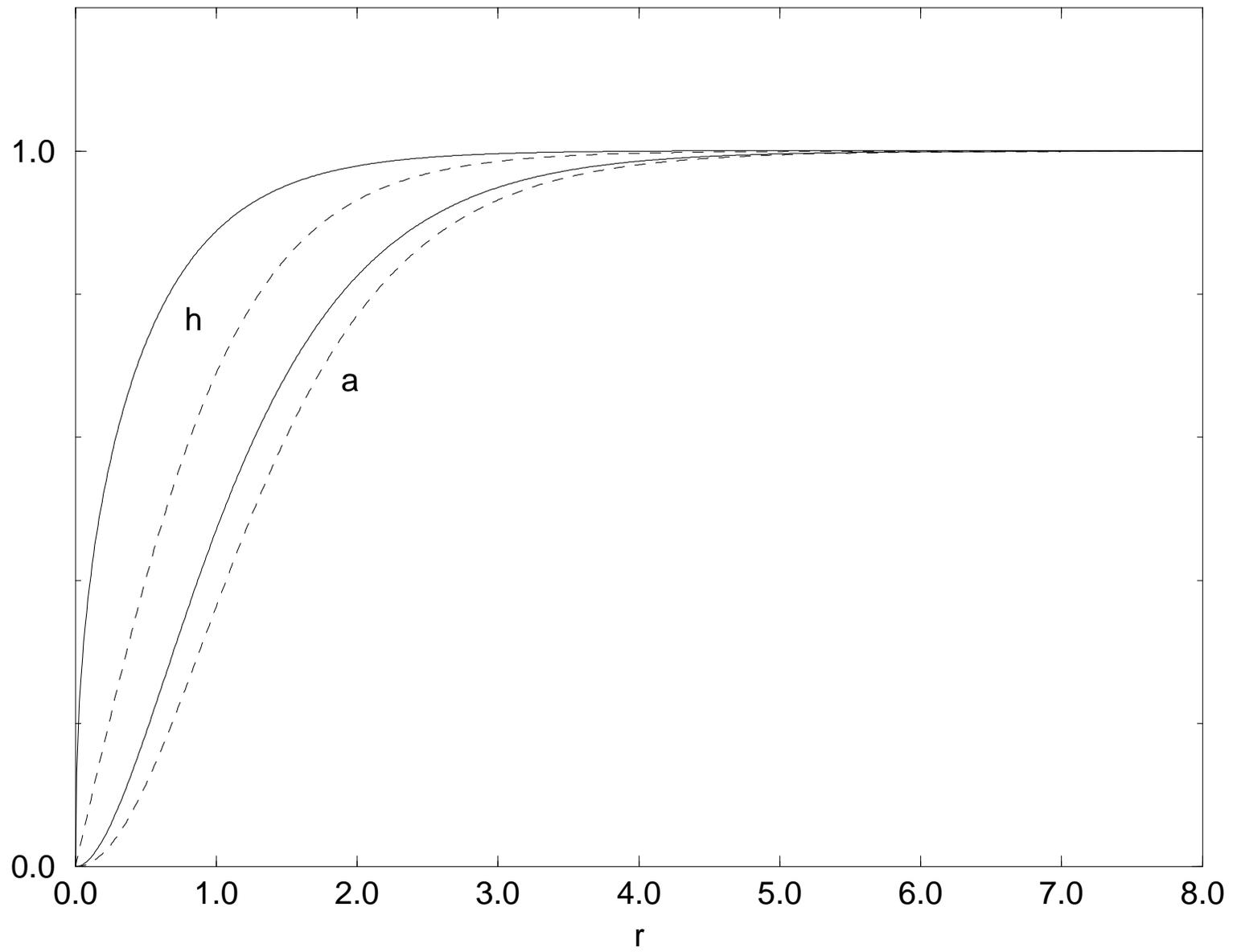
\end{document}